\documentclass{article}

\usepackage{algorithm2e}
\usepackage{amsmath}
\usepackage{graphicx}

\title{Focused Proofreading: Efficiently Extracting Connectomes from Segmented EM Images}

\author{Stephen M Plaza}

\begin{document}
\maketitle

\begin{abstract}
Identifying complex neural circuitry from electron microscopic (EM) images 
may help unlock the mysteries of the brain.  However, identifying this
circuitry requires time-consuming,
manual tracing (proofreading) due to the
size and intricacy of these image datasets, thus limiting state-of-the-art analysis to very small brain regions.  Potential avenues to improve scalability include automatic image segmentation and crowd sourcing, but current efforts have had limited success.  In this  paper, we propose a new strategy, focused proofreading, that works with automatic segmentation and aims to limit proofreading to the regions of a dataset that are most impactful to the resulting circuit.  We then introduce a novel workflow, which exploits biological information such as synapses, and apply it to a large dataset in the fly optic lobe.  With our
techniques, we achieve significant tracing speedups of 3-5x without
sacrificing the quality of the resulting circuit.  Furthermore, our methodology
makes the task of proofreading much more accessible and hence potentially enhances 
the effectiveness of crowd sourcing.

\end{abstract}

\section{Introduction}
EM reconstruction is the process of extracting a connectome from an EM
dataset.  A structural connectome derivable from EM data typically consists
of neurons and their connections/synapses.  To decipher the
intricacy of neuronal 
structures in a brain, the imaging is at nanometer resolution generating vast
amounts of data to be analyzed.  Because of this, reconstruction is very
time consuming and significant advances are needed to handle larger volumes \cite{Plaza14}.

Two main approaches exist for reconstructing
connectomes from an EM dataset: manual skeletonization and refinement of automatic
segmentation.  Skeletonization requires a proofreader to manually trace the shape of the cell \cite{CATMAID, Winfried13}.  CATMAID \cite{CATMAID} achieves some
scalability success by making collaborative, web-based tracing very accessible to interested, well-trained biologists.  In \cite{Winfried13}, skeletonization is accomplished
through a consensus of, generally, less well-trained students.  Segmentation-driven tracing has been successfully deployed in a partial reconstructions of the fly optic lobe \cite{Nature13} and mouse retina \cite{Sebastian14}.  Reconstruction is achieved by merging and splitting incorrectly segments, which still results in effort.

\begin{figure}
\centering
\includegraphics[width=1.0\textwidth]{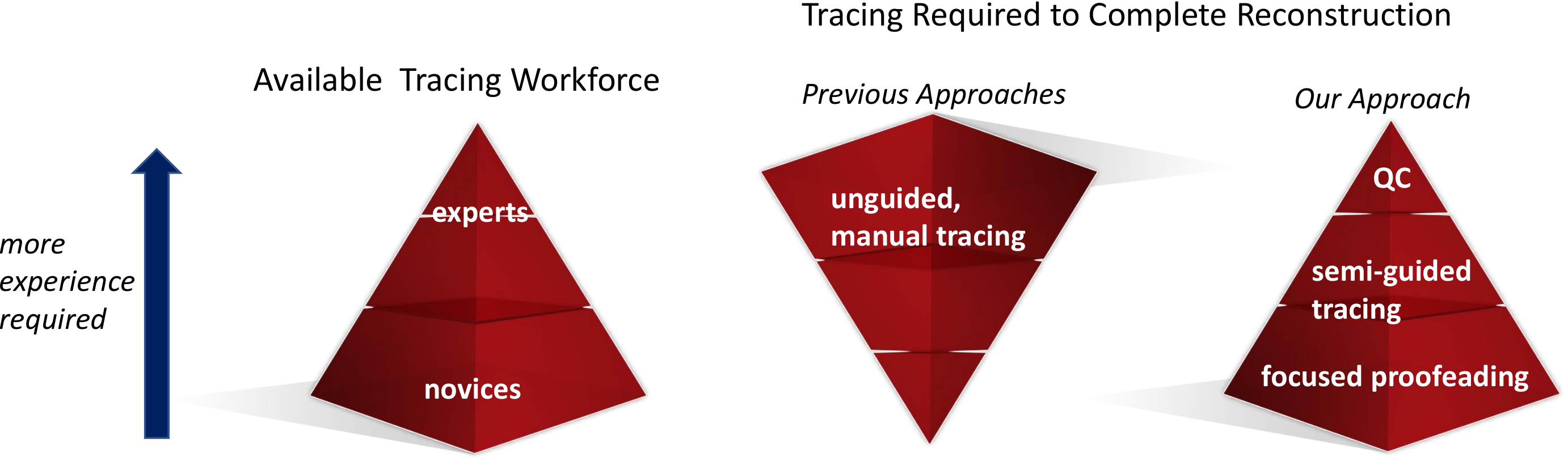}
\caption{\label{fig:pyramid} {\bf The manual tracing dilemma.}  Despite a potentially large available workforce, the difficulty
of manual tracing in traditional approaches often requires
more expertise than is feasible.  Some approaches that use consensus skeletonization \cite{Winfried13} attempt to
expand the available workforce by averaging the often less-than-optimal tracings of multiple proofreaders.  However, a) this requires even more proofreaders, b) there is still a verification bottleneck, and c) important features could be averaged out.}
\vspace{-1mm}
\end{figure}

In the ideal case, automatic segmentation would produce a perfect connectome.  While recent advances in EM segmentation \cite{Huang14, jni13, Parag14, Andres, Funke12} (particularly in isotropic FIB-SEM datasets \cite{fib}) produce very good results, the segmentations are still far from perfect.  This is due, in part, to the sensitivity of the connectome to small segmentation errors.  For instance, a small false merger in a small part of the image can cause two separate neurons to be merged together greatly affecting the connectivity map.  Even with near-perfect segmentation, extensive verification is likely to be required.  Better segmentation, alone, will not solve scalability.

Crowd-sourcing has been pursued in different ways \cite{CATMAID, Winfried13, Sebastian14} as a potential solution.  However, these strategies are fundamentally unscalable.  Figure \ref{fig:pyramid} illustrates the dilemma of crowd-sourcing EM reconstruction.  Traditionally, EM tracing requires a high-level of expertise requiring weeks of training (or more), unreasonable for a general crowd-source community.
CATMAID \cite{CATMAID} tries to expand the expert base through its accessibility but is not meant for beginner tracers.  Consensus tracing, as in \cite{Winfried13} can access a wider pool but requires even more
proofreaders to account for errors.  Also, an averaged result could lead to a sub-optimal connectome or require extensive expert verification.  The approach in \cite{Sebastian14} attempts to make proofreading accessible to the novice community.  Despite tremendous involvement from the community, the efforts were primarily used for validation, and the reconstruction still required a group of trained proofreaders.

To address these scalability challenges, we propose {\em focused proofreading}.
Focused proofreading is a segmentation-driven proofreading that attempts to
discern the regions of the segmentation that are both relevant to the connectome and least-likely to
be correct.  In the process, it distills the task of proofreading to a more
digestible series of yes/no decisions.  By redefining proofreading, we hope to expand the base of potential proofreaders
as shown in Figure \ref{fig:pyramid}.  Our work has some similarities
to the uncertainty-driven proofreading suggested in \cite{Plaza12}.  However, we propose a more practical approach that uses efficiently computed local constraints to guide proofreading rather than a global strategy.  Furthermore, we exploit synapse information and other biological priors to greatly enhance proofreading and the quality of the final reconstruction.

To effectively guide proofreading, we introduce several metrics aimed
at understanding what it means to complete a reconstruction.  In particular,
we propose a new, connectivity-based similarity metric to assess the quality
of a segmentation and to subsequently guide efforts.  Previous segmentation
efforts were primarily concerned with producing highly similar segmentation at
only the voxel level.  This paper attempts to add more biological relevance to
this analysis.  Focused proofreading leads to
significant improvements compared to random or other proofreading strategies.

\begin{figure}
\centering
\includegraphics[width=1.0\textwidth]{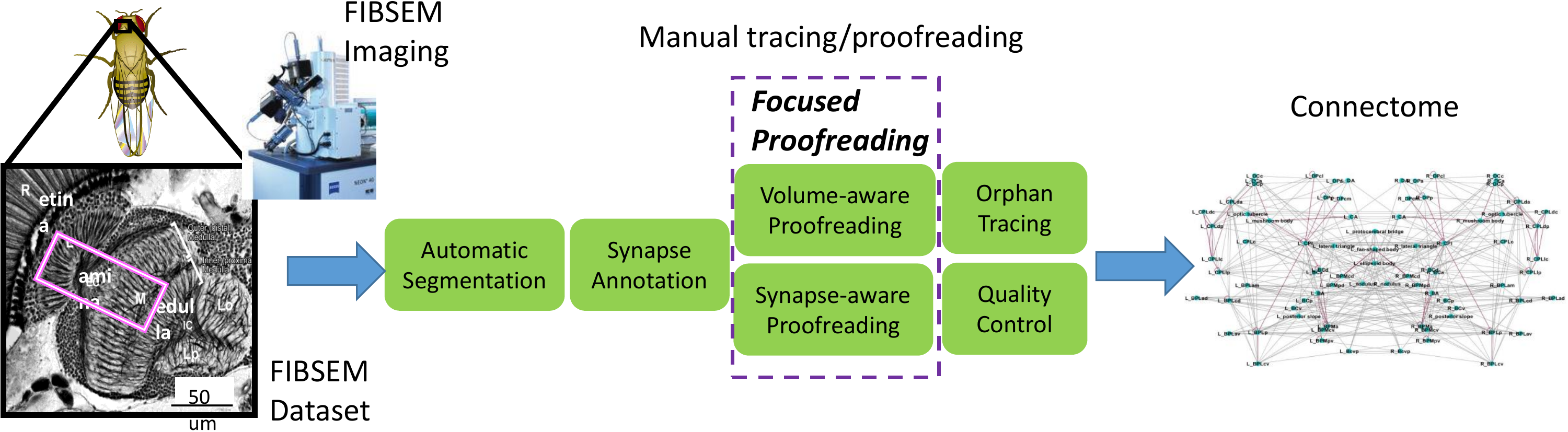}
\caption{\label{fig:reconoverview} {\bf Reconstruction pipeline to produce a connectome.}  This paper introduces focused proofreading methodology and considers using synapse annotations to guide proofreading.}
\vspace{-1mm}
\end{figure}

In this paper, we apply our focused proofreading techniques to reconstruct
seven medulla columns in the Drosophila optic lobe.  Through the combination
of improved imaging \cite{fib} and the methodology introduced here, we finished more in less time than the one-column reconstruction in \cite{Nature13}.  Our best estimates indicate 3-5x speedup in reconstruction with improved accuracy.  The general flow is depicted in Figure \ref{fig:reconoverview}.

The paper is organized as follows.  We first investigate similarity metrics and provide more context on the challenges of segmentation-driven reconstruction.  Then we introduce metrics to help better define what is meant by a good, or complete, reconstruction.  We next introduce the focused proofreading algorithm and discuss its practical deployment.  Results are given for
a ground truth subvolume, and statistics are provided from the entire reconstruction.

\section{Background: Similarity Metrics}
Since the goal of this paper is to introduce techniques to best guide manual
annotation, we first examine what it means for manual annotation (and similarly
automatic segmentation) to be close to the correct result.  We can consider the general
case of assessing the quality of a labeled image volume compared to a gold-standard,
or so-called ground truth.
In production (test) workflows, such ground truth is obviously unavailable
in general.  However, groundtruth on small regions of image data
can act as a proxy for algorithms or methodology in similar regions.

The differences between a segmented label volume (S) and ground truth (G) can be
quantified by considering the variation of information (VI) \cite{vi}:

\begin{equation}
\label{eq:vi}
VI(S,G) = H(S|G) + H(G|S)
\end{equation}

where H is the entropy function.  The first term, $H(S|G)$ gives the information
of the underlying segmentation compared to ground truth and indicates over-segmentation.
Likewise, $H(G|S)$ indicates under-segmentation.  $0$ information means high similarity.  Plotting both terms will produce
a precision/recall-like curve.  While other similarity
metrics are used to compare label volumes, such as Rand Index \cite{rand,adjustedrand}
and Warping Index \cite{warping}, VI is both simple to compute and has interpretability
advantages as highlighted by the authors in \cite{jni13}.  As with computing
the Rand index, careful use of hash-map datastructures enable VI to be computed
in time roughly linear to the dataset size.  While this paper will emphasize VI,
other measures, like Rand Index, could be adapted.

Typically, the metric is applied
to partitions over the voxel space where all pixels are considered.  As
suggested in \cite{jni13}, we can decompose the over-segmentation
components of VI by:

\begin{equation}
\label{eq:vioverseg}
H(S|G) = -\sum_g{P(g)H(S|G=g)}
\end{equation}

This allows introspection as to which ground truth label $g$ observes
the greatest over-segmentation.  Note that the impact of a given
labeling with respect to a ground truth label $g$ is weighted by $P(g)$.
Consequently, big incorrect bodies contribute significantly more than
small bodies.  This detail will be explored later.  A similar analysis
can be done for $H(G|S)$.

\section{Background: Segmentation-driven EM Reconstruction}
As imaging and machine learning algorithms have improved,
annotating datasets with the help of automatic segmentation is
more prevalent \cite{Nature13, Sebastian14}.  As explained
in \cite{Plaza14}, the ability to extend analyses to increasingly
large datasets depends on effectively exploiting what the
computer can extract automatically.  The primary mechanism
involves creating an initial segmentation.  This segmentation is
then revised often by merging label regions that were erroneously
split.
In practice, it is much
easier to refine an oversegmented label volume than an under-segmented
volume.

A problem arises when reconstructing connectomes in this manner.
Which segments should be examined?  Where should an annotator focus
his/her attention?  If the initial segmentation
is poor, everything must be examined and extensive effort must be spent correcting it, as captured
by the so-called {\em nuisance metric}.  Under these
circumstances, it is hard to see how this approach is more scalable than
manually annotating datasets using skeletonization \cite{CATMAID, Winfried13}.
In fact, skeletonization could be faster since the
annotator can ignore parts of the dataset 
irrelevant to a specific connection pathway.

If the initial segmentation is good and few corrections are needed, any segmentation-driven strategy is likely superior to skeletonization.
However, even with 100 percent correct segmentation (zero nuisance), verifying
that it is really 100 percent correct on a large dataset could still require thorough inspection by
an army of annotators. To improve this, the annotators could
be focused to examine only areas where the segmentation likely
erred.  The concept of optimizing which regions to
examine was first considered in \cite{Plaza12}.  That work defines metrics for determining the correctness of segmentation in the absence of ground truth.

\section{Proofreading Completeness Measures}
\begin{figure}
\centering
\includegraphics[width=1.0\textwidth]{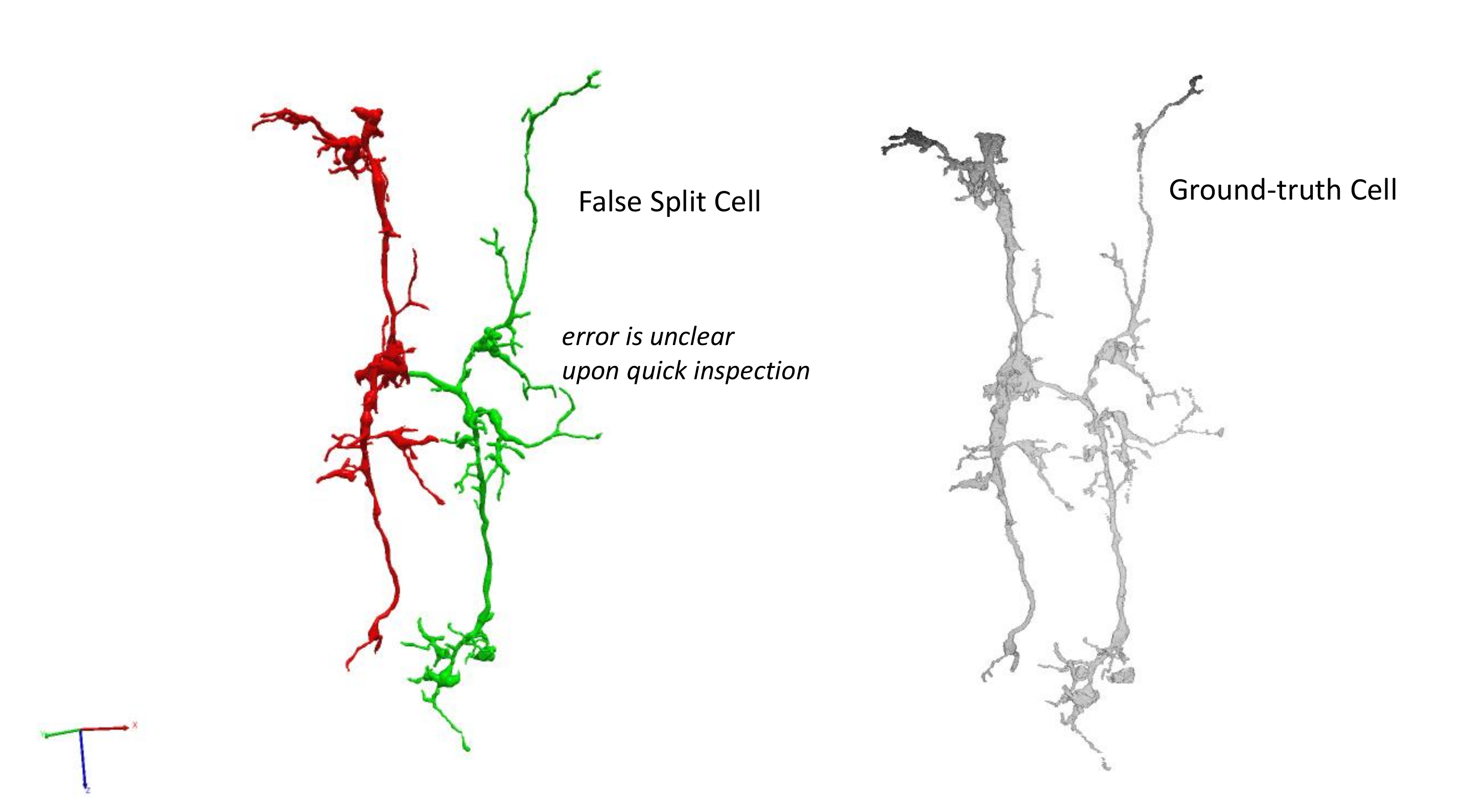}
\caption{\label{fig:hardtrace} Here a ground truth cell is falsely split at a small branch that results into two segments that look like individual neurons.  Segmentation errors like this (even if they are very rare) are not always evident and may require exhaustive
verification without adequate proofreading guidance.}
\vspace{-1mm}
\end{figure}

In practice, the segmentation will significantly deviate
from ground truth.
Furthermore, any strategy that selectively examines the
segmentation, unless it is an oracle, will likely miss errors.  Consider
Figure \ref{fig:hardtrace}, where it appears that there are two complete
neuron shapes.  It is only through detailed inspection can one find the small
connection that joins these two neuron segments together.
But as we noted previously, examining everything is intractable.
Therefore, we attempt to better quantify errors like this one
and other less important, inconsequential errors and their relevance to the connectome.

The barebone definition of a connectome is a
graph whereby each node represents a disjoint
neuron and the edges represent synaptic connections.  However,
the actual shape of the neuron is also useful for identifying
like nodes and comparing neurons between datasets.
Furthermore, volumetric accuracy, as well as actual synaptic locations
could be useful for circuit modeling.
Additional detail might be sought depending on the application (e.g.,
distribution of synaptic vesicles, mitochondria, etc).
The metrics defined in this paper will consider a connectome
as one that contains most of the neurons and most of their volume
and connections.  \footnote{If the readers' definition of a connectome differs, the subsequent
goals metrics should be modified accordingly.}
Because pixel-perfect accuracy is unnecessary to
identify neurons and most neuronal pathways contain multiple connections,
we can create metrics that tolerate some errors.  We will
adjust the input to the VI metric so that some {\em errors} do
not penalize the similarity between ground truth and label volume.
The following considers the problem of over-segmentation.  The reverse
is assumed to happen infrequently by construction.

\subsection{Volumetric Completeness}
\begin{figure}
\centering
\includegraphics[width=1.0\textwidth]{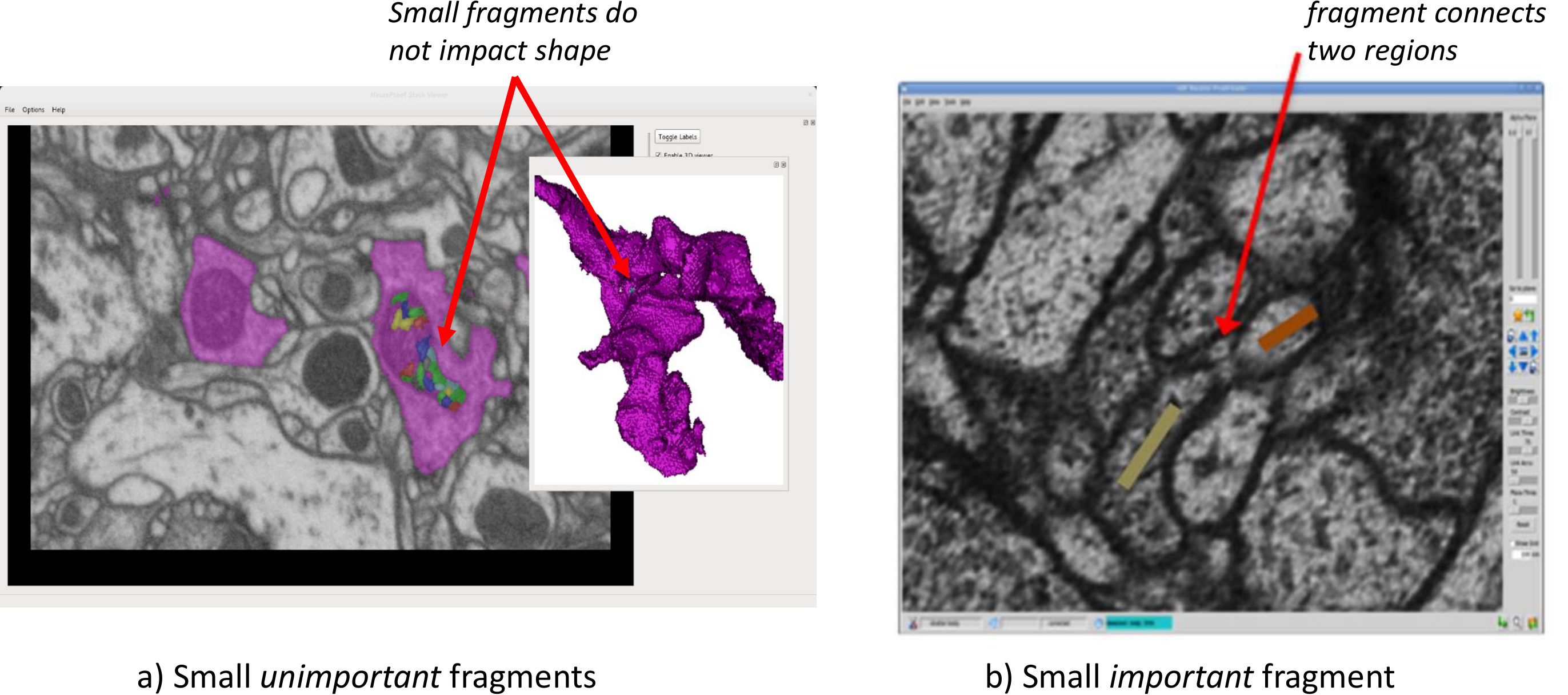}
\caption{\label{fig:smallbodies} a) Shows small incorrectly assigned segments that do not impact the shape (and the resulting skeleton) of the neuron.  b) Shows a small segment that connects two different larger regions.}
\vspace{-1mm}
\end{figure}

This section will focus on the completeness of a connectome
with respect to the neurons' volume and shape.
Given an over-segmentation with a set of labels $S$,
a complete volumetric proofreading involves assigning (merging)
each $s$ to some final $g$
(typically a connected component).
Each $g$ constitutes a distinct neuron.
However, if only the shape is desired, a skeletonized representation
can often ignore several $s$. 
Figure \ref{fig:smallbodies}a shows an example of an
oversegmented neuron where various
``unimportant'' segments are highlighted.  Note, though, that a small $s$
can be important as in Figure \ref{fig:smallbodies}b.  
We can modify the VI metric in Equation \ref{eq:vioverseg} by
considering
only the errors that impact shape.  

\begin{equation}
\label{eq:skel}
H(S|G)_{skeleton} = -\sum_{g_s}{[\frac{|g_s|}{|G_s|}\sum_s{-\frac{match(s_s,g_s)}{|g_s|}log_2(\frac{match(s_s,g_s)}{|g_s|})}]}
\end{equation}

where $g_s$ and $s_s$ refer to the set of points (or lines)
that define the skeletons of $g$ and $s$.
$match$ could then define some correspondence function
between $s_s$ and $g_s$.
For this paper, we still desire volumetric accuracy.  We redefine
$g_s$ and $s_s$ to $g_{vol}$ and $s_{vol}$ which now denote the sets of voxels
associated with each label, as in the traditional usage of VI.  $match$ is then the intersection between
these voxel sets.

\begin{equation}
H(S|G)_{volume} = -\sum_{g_{vol}}{[\frac{|g_{vol}|}{|G_{vol}|}\sum_{s_{vol}}{-\frac{|s_{vol} \wedge g_{vol}|}{|g_{vol}|}log_2(\frac{|s_{vol} \wedge g_{vol}|}{|g_{vol}|})}]}
\end{equation}

Analogous to ignoring segments
irrelevant to the shape in Equation \ref{eq:skel}, unimportant $s$
can be de-emphasized by removing voxels from $s$ and $g$ that are
close to the surface of $g$ through a label erosion image operation.
In this manner, noisy boundaries that have little impact on the volume
or correspondence between $s$ and $g$ are ignored.

The preceding formulae more accurately reflect the true similarity
between $G$ and $S$.  But volumetric differences will still exist.
We attempt to define a threshold for
an acceptable number of differences (completeness) with the following:

\begin{equation}
\label{eq:compl}
Completeness(S,G): \lambda < \sum_g I[k > P(g)(H|G_g)]
\end{equation}

where $P(g)$ is the importance or frequency of $g$ in $G$ and $I$ is the indicator function.  In other words, automatic segmentation and subsequent proofreading refinement
should be considered complete when there are less than $\lambda$ bodies
incorrect with threshold $k$.  Unlike just setting up a global threshold for
Equation \ref{eq:skel}, this formula attempts to decompose the problem into
something more biologically interpretable, a per body constraint.
It also distinguishes between scenarios where several small differences (all
insignificant to every body of importance) exceed some global threshold.

Equation \ref{eq:compl} leads to two questions: 1) what is a good $k$
and 2) what happens when $G$ is unavailable?  The answer to the second
question is mainly considered in the next section.

\begin{figure}
\centering
\includegraphics[width=0.8\textwidth]{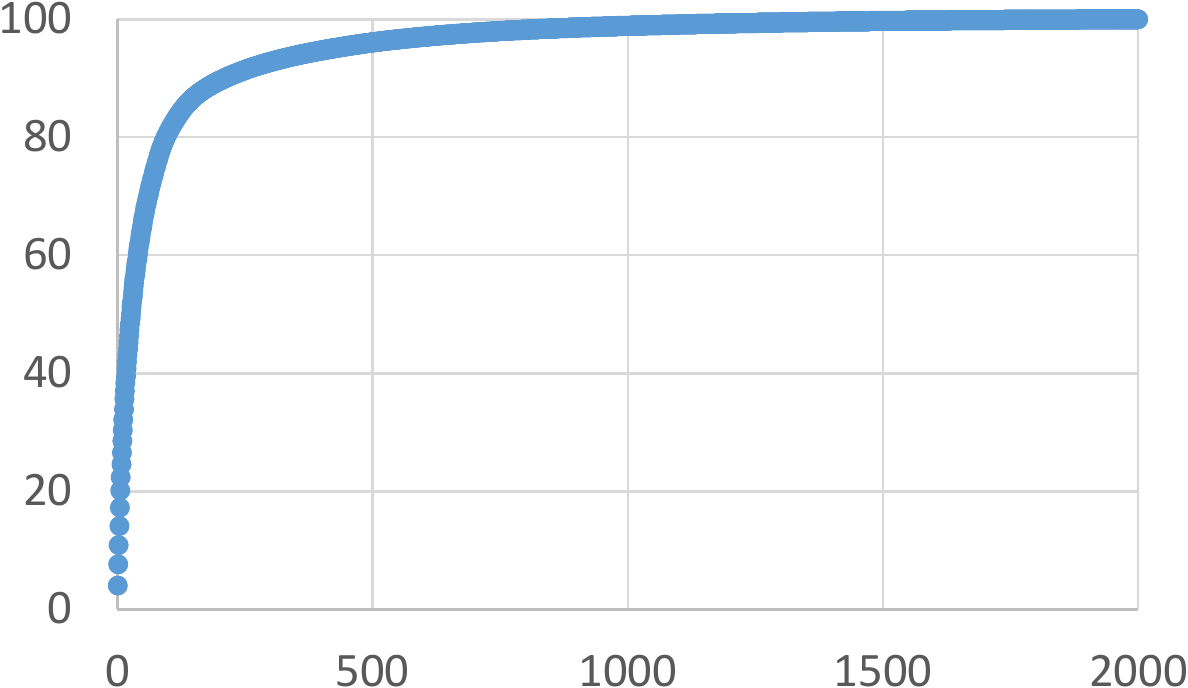}
\caption{\label{fig:gtdist} Plot that shows that a small number of bodies (a few hundred) contribute to over 90\% of the volume for a typical example.  The long tail is due to bodies near the edge of this ground truth volume and small inaccuracies or untraceable processes.  We can estimate completeness by ensuring we see a similar distribution of large neurons in our proofread .}
\vspace{-1mm}
\end{figure}

We can better understand $k$ and $G$ by analyzing the distribution of expected neuron sizes for a representative region as in Figure \ref{fig:gtdist}.
Notice that a small number of bodies constitute a large fraction of
the volume.  The small bodies consist of 
untraceable, {\em orphan} processes, or
neuronal arbors clipped by the
edge of the volume.  Any chosen threshold should ensure that
a given region would have several large or non-orphan
neurons consistent with the distribution in Figure \ref{fig:gtdist}.
$k$ can then be derived where a) segments larger than $|s_{th}|$ are deemed important and where b)
we expect the average size of $g$ to be $|\bar{g}|$:

\begin{equation}
\label{eq:thres}
k = -\frac{|s_{th}|}{|n|}log_2(\frac{|s_{th}|}{|\bar{g}|}) - \frac{|\bar{g}|-|s_{th}|}{|n|}log_2(\frac{|\bar{g}|-|s_{th}|}{|\bar{g}|})
\end{equation}

where $|n|$ is the size of the whole volume and can be factored away.  This equation expands the $H(S|G)$ for a given $\bar{g}$ defining
an undesirable two-way partitioning of the body.  This formula
is somewhat academic since a given body can be partitioned
into many pieces.  However, this formulation is relevant as a stopping condition for focused proofreading as defined in the next section.  In practice,
we just note that  small orphan (disconnected) segments are biologically implausible and should be ignored below a certain size $|s_{th}|$. 

\subsection{Synaptic Completeness}
\begin{figure}
\centering
\includegraphics[width=1.0\textwidth]{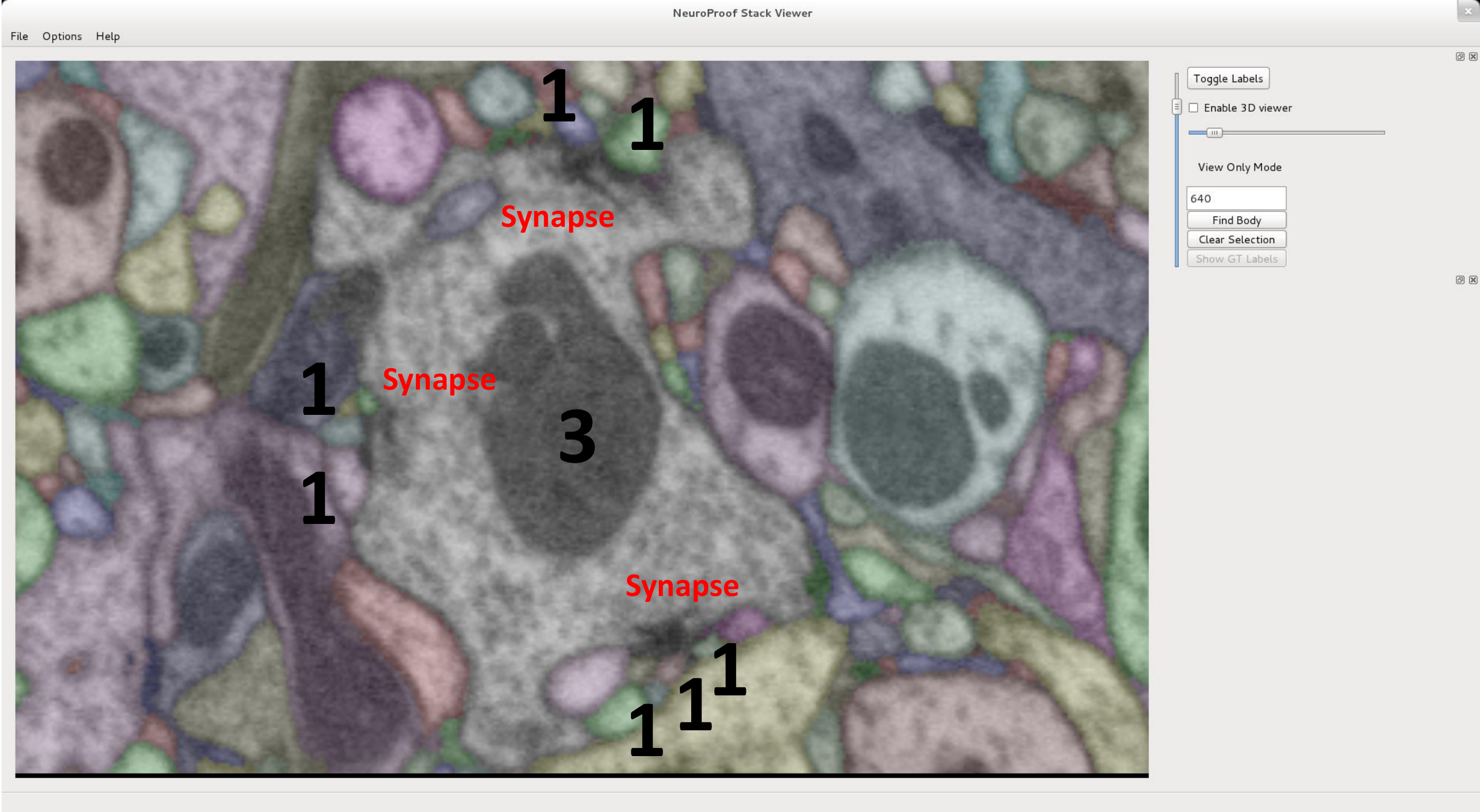}
\caption{\label{fig:synapsevi} Synapse VI similarity metric.  Similarity is computed over a segmentation by examining the number of synapse annotations in each segment.  The middle segment has weight $3$ since it has three pre-synaptic regions.  The outputs to these synapses are each assigned a weight of $1$.  The importance of the bodies is not determined by their size in voxels.}
\vspace{-1mm}
\end{figure}

We introduce a new metric for analyzing the quality of segmentation
that considers the connectivity.  The concepts and motivation
are analogous to those presented in the previous section.  We define
the synaptic $H(S|G)_{synapse}$ as:

\begin{equation}
\label{eqn:visyn}
H(S|G)_{synapse} = -\sum_{g_{syn}}{[\frac{|g_{syn}|}{|G_{syn}|}\sum_{s_{syn}}{-\frac{|s_{syn} \wedge g_{syn}|}{|g_{syn}|}log_2(\frac{|s_{syn} \wedge g_{syn}|}{|g_{syn}|})}]}
\end{equation}

where $s_{syn}$ and $g_{syn}$ are just a subset of $s$ and $g$
that contain synaptic annotations.  Figure \ref{fig:synapsevi} above shows
a body with three presynaptic regions and multiple post synaptic bodies
as partners.  For example, the middle segment $a$ would have $|s_{syn}| = 3$ while $|s_{vol}|$ would be the number of voxels in $s$.  Equation
\ref{eqn:visyn} does not distinguish between pre and post-synaptic
regions.  $H(G|S)_{synapse}$ can be analogously defined, and likewise
$VI_{syn}(S,G)$.  While previous segmentation efforts focus exclusively on volume, synapse VI is probably the most relevant single measure for connectome similarity.

Since synaptic completeness is similar to volume, we just
note the following.  First, unlike the volume measure
there is little ambiguity about the relevance of this measure.  Every difference
corresponds to a concrete change in the connectome.  Despite this,
there may be considerable flexibility when defining completeness.
In examined Drosophila nervous systems, there is often considerable
redundancy of connections in the strongest pathways.  Coupling that
with inherent variability and plasticity, we can set a threshold
to a level acceptable for subsequent analysis.  As with volumetric
completeness, we also note that a biologically correct connectome cannot have synapses in small, orphan segments.

\section{Uncertainty-driven Focused Proofreading}
The previous section defines what is ``good enough''.  This section
defines how to get there.  We consider the case where a proofreader
is given $S$ and must revise it to $S'$, so that it is reasonably
close to $G$.  However, for simplicity of analysis, this proofreader
is restricted to merge-only operations (we consider
workflows handling splits in the next section).  More specifically, we consider
the proofreading problem as an assignment of {\em yes} or {\em no}
for edges in $S$, where an edge connects two neighboring $s$.

We can define an optimal similarity after $m$ decisions as:

\begin{equation}
sim(m) = min_{S_{\pi(m)}}(H(S_{\pi(m)}|G))  
\end{equation}

where $\pi(n)$ denotes an optimal ordering of $m$ ${yes, no}$ decisions.
Since we are starting from an oversegmented $S$, $\pi(m)$
consists of an ordering of only $m$ no (merge) decisions.
We define $m^*$ as the optimal number of decisions to
achieve $completeness(S_{\pi(m^*)},G)$.

We do not attempt to solve $\pi(m)$ optimally.  Instead we favor
greedy-based orderings that have the greatest impact on $H(S|G)$.  However, simple, greedy-based approaches have two problems.
First, explicit $G$ is unavailable for measuring impact. 
Second, as Figure \ref{fig:smallbodies}b illustrates, sometimes two large $s$ are disconnected requiring a {\em smaller} decision to be made first.
We address these concerns in the following two parts.

\subsection{Uncertainty-determined Ground truth}
Assuming a greedy-based decision ordering, we seek an impact
measure for ranking an edge.  The impact of the 
edge between segments $a_i$ and $a_j$ is analogous to the threshold equation defined in
Equation \ref{eq:thres}:

\begin{equation}
\label{eq:impact}
Impact(e_{(i,j)}) = -|a_i|log_2(\frac{|a_i|}{|a_i|+|a_j|})-|a_j|log_2(\frac{|a_j|}{|a_i|+|a_j|})
\end{equation}

$|a_i|$, and $|a_j|$ could represent the number of voxels or synaptic annotations in $a_i$, and $a_j$ respectively.  Note that we removed
the normalizing $|n|$ in Equation \ref{eq:thres} as it does not
affect the ordering.

\begin{figure}
\centering
\includegraphics[width=1.0\textwidth]{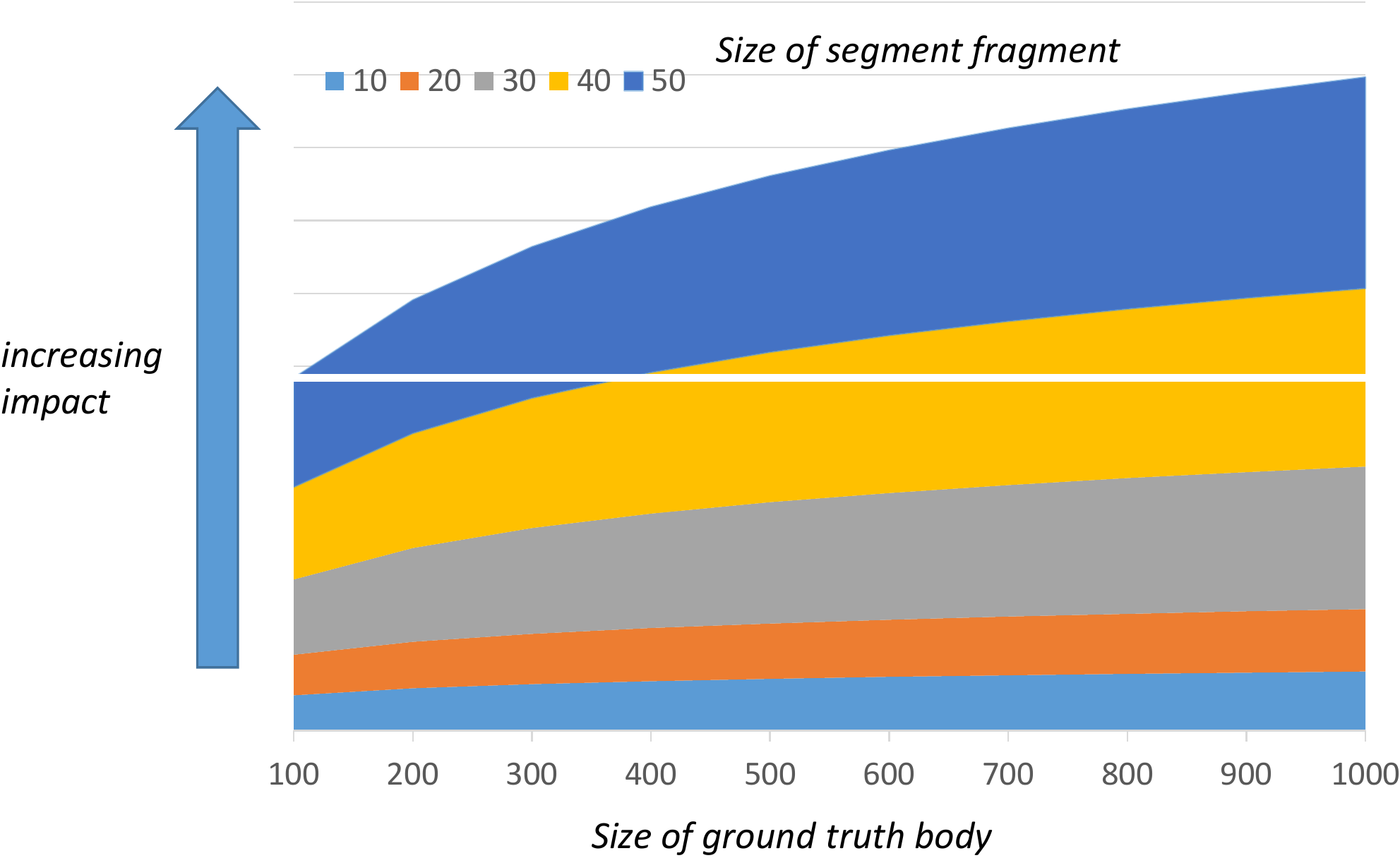}
\caption{\label{fig:impacteq} Impact as a function of the total body size and the smaller partition as in Equation \ref{eq:impact} but with $g_{spec}$ substituted for $a_i + a_j$.  The plot shows that larger ground truth bodies, logically, are more impactful.  Less intuitively, the horizontal line of equal impact indicates that a smaller fragment in a larger neuron is more impactful than a larger fragment in smaller neuron.}
\vspace{-1mm}
\end{figure}

We can view $a_i \wedge a_j$ as being a speculative ground truth, $g_{spec_{ij}}$.
Figure \ref{fig:impacteq} shows a plot of
Equation \ref{eq:impact}
where $g_{spec_{ij}}$ is substituted and $a$ is meant to be the smaller of $a_i$ and $a_j$.  It indicates that impact increases as both $g_{spec_{ij}}$ and $a$ increase.  Notably, a given edge can be more
impactful than another edge if $g_{spec_{ij}}$ is larger, even if
the ratio between $a$ and $g_{spec_{ij}}$ is smaller.  Is this
desirable?  First, this impact measure better optimizes the VI
similarity measure.  Second, it suggestions that bigger, more
complete bodies, should be examined first.  For example, if $|g_{spec_{ij}}| > |g_{spec_{kl}}|$ and $a_i$ and $a_k$ represent
a single missing synapse, the impact measure tries to fix errors
in the, likely, more-complete $g_{spec{ij}}$ first.

Examining the most impactful true edges is undesirable
since the quality of $S$ will not improve.
The risk of each edge needs to be quantified.

\begin{equation}
\label{eq:risk}
Risk(e_{(i,j)}) = P(\neg e_{(i,j)}) Impact(e_{(i,j)})
\end{equation}

The riskiest edges define the edges that will likely have the greatest
impact on this segmentation.  There is some similarity between this
and the impact measures defined in \cite{Plaza12}.  However, that
work focused on measures that reduced the uncertainty in a segmentation,
so that additional automatic segmentation could be performed.  This work
does not leverage additional segmentation and tries to minimize the likelihood that false mergers will occur.  This formulation is also
more computationally economical as it does not require a global
model of uncertainty.  In other words, our approach is more practically
deployable.

To determine $P(\neg e_{(i,j)})$, we first train a classifier on the
edges of an oversegmented volume.  The resulting prediction determines
the confidence in the edge.  This classifier is trained similarly to
those discussed in \cite{Parag14, jni13}. 

We chose the random forest classifier since it performs well, while
being fast and easy to deploy.  Its predicted uncertainties also conform
reasonably well to the actual ground truth as shown in the experiments.

\subsection{Focused Proofreading Algortihm}

We define {\em focused proofreading} as the examination
of a subset edges where $Risk(e_{(i,j)}) > k$, where $k$
is determined by Equation \ref{eq:thres}.
The greedy-based strategy introduced previously is flawed
since two labels $s_i$ and $s_j$ might belong to the same neuron
but have no direct edge (as seen in Figure \ref{fig:smallbodies}b).
As in \cite{Plaza12}, we avoid this problem by considering the probability that 
a set of edges connect $s_i$ and $s_j$:

\begin{equation}
P(\neg E_{(i,j)}) = \Pi_{\neg e_{k,l} \in E_{i,j}}P(e_{k,l})
\end{equation}

where $P(\neg E{(i,j)})$ is the probability of a path existing
between $s_i$ and $s_j$.  By finding potential paths between
pairs of labels, the greedy-based strategy of choosing the
riskiest pair can have more global awareness.

For certain over-segmentations, examining all paths
where $Risk(E_{(i,j)}) > k$ will still not produce a good reconstruction, even if the predictions are exact.  An extreme is example would be oversegmenting a given body into individual
voxels.  In this case, no two segments are important; the collection is important.  While it might be possible to define a strategy that determines whether a set of segments has affinity, we note that such circumstances should be rare in a reasonable segmentation and its existence would likely strain the quality of any uncertainty estimation.  We can define a reasonable segmentation with the following two conditions:

\begin{equation}
\sum_g I((\sum_{g_a: (|g_a| < x)}|g_a|) < |s_{th}|) < \alpha
\end{equation}
\begin{equation}
\sum_g I(\forall_{i,j \in g | (|s_i|,|s_j| < |s_{th}|} |E_{(i,j)}| < \beta) < \gamma
\end{equation}

The first equation is a constraint that says that a certain number of neurons ($\alpha$) should be covered by segments of at least $x$ size to reach the desirable threshold $|s_{th}|$.  In the degenerate case, $x$ could be chosen to be really small, which would always satisfy this condition.  But this may increase the amount of spurious proofreading required and might violate the second equation.  The second equation says that these important segments of size $|s_{th}|$ must be connected within $\beta$ hops.
Ideally, the $P(E_{i,j})$
should also be within some threshold as well.  When using $VI_{synapse}$ we often set the threshold to just one annotation trivially satisfying the first condition.  When using $VI_{volume}$, not all pixels are important and we set the threshold in a manner that gives us good coverage.  We show this result in the experiment section.

\vspace{2mm}
\noindent
{\bf Algorithm}
\vspace{1mm}

We organize the previous thoughts and now present an algorithm for proofreading an oversegmentation.

\begin{algorithm}
\SetKwFunction{findNeighbors}{findNeighbors}
\SetKwFunction{risk}{risk}
\SetKwFunction{decide}{decide}
\KwIn{Segmentation: S, Threshold: k}
\KwOut{Proofread Segmentation: S}
\ForEach {$s_b$ $\in$ S'}{
	$S_E$ = \findNeighbors{$s_b$}\;
	\ForEach {$s_a \in S_E$}{
    	\If{\risk{$E(s_a,s_b)$} $>$ k}{
        	result = \decide{$E(s_a,s_b)$} \;
        	S $\leftarrow$ result \;
            $S_E$ = \findNeighbors{$s_b$}\;
        }
    }
}
\label{alg:fp}
\caption{Focused proofreading algorithm.}
\end{algorithm}

The algorithm uses a threshold $k$ determined heuristically.  Since completeness is desired, it is not as relevant to choose the best order to examine
edges.  It suffices to examine all edges within a threshold.  (We can effectively show the quality of the algorithm as a function of decisions by successively lowering this threshold or by constraining the algorithm to ignore low $P(\neg E)$ .)  The algorithm starts by iterating through all segments considering the {\em largest} first.  Then, all potential segments connected to this body (within some uncertainty threshold) are determined through
function {\tt findNeighbors}.  The proofreader is given edges along the riskiest path in {\tt decide}.  After each decision, the graph  and list of candidate edges are updated.

If the segmentation is reasonable, the focused proofreading can still perform poorly if the uncertainties are poor.  If the uncertainties favor false merging, the algorithm will lead to inefficiency, as many true edges will be examined.  If the uncertainties favor false splitting, errors will occur in the final segmentation.  This affect is mitigated slightly since very impactful decisions can still be examined even if the true edge probability is high.  The next section discusses how this algorithm is deployed in practice.

\section{Proposed Workflow}
We deployed the algorithm described previously to reconstruct
the neuronal pathways in the medulla columns of the Drosophila
optic lobe, which contained hundreds of partial neurons and
several hundred thousand synaptic connections.  In practice, there are many challenges to reconstruction.  1) The initial segmentation will falsely merge some regions.  2) Focused proofreading will miss some important areas.  3) Proofreaders will make errors.

\begin{figure}
\centering
\includegraphics[width=1.0\textwidth]{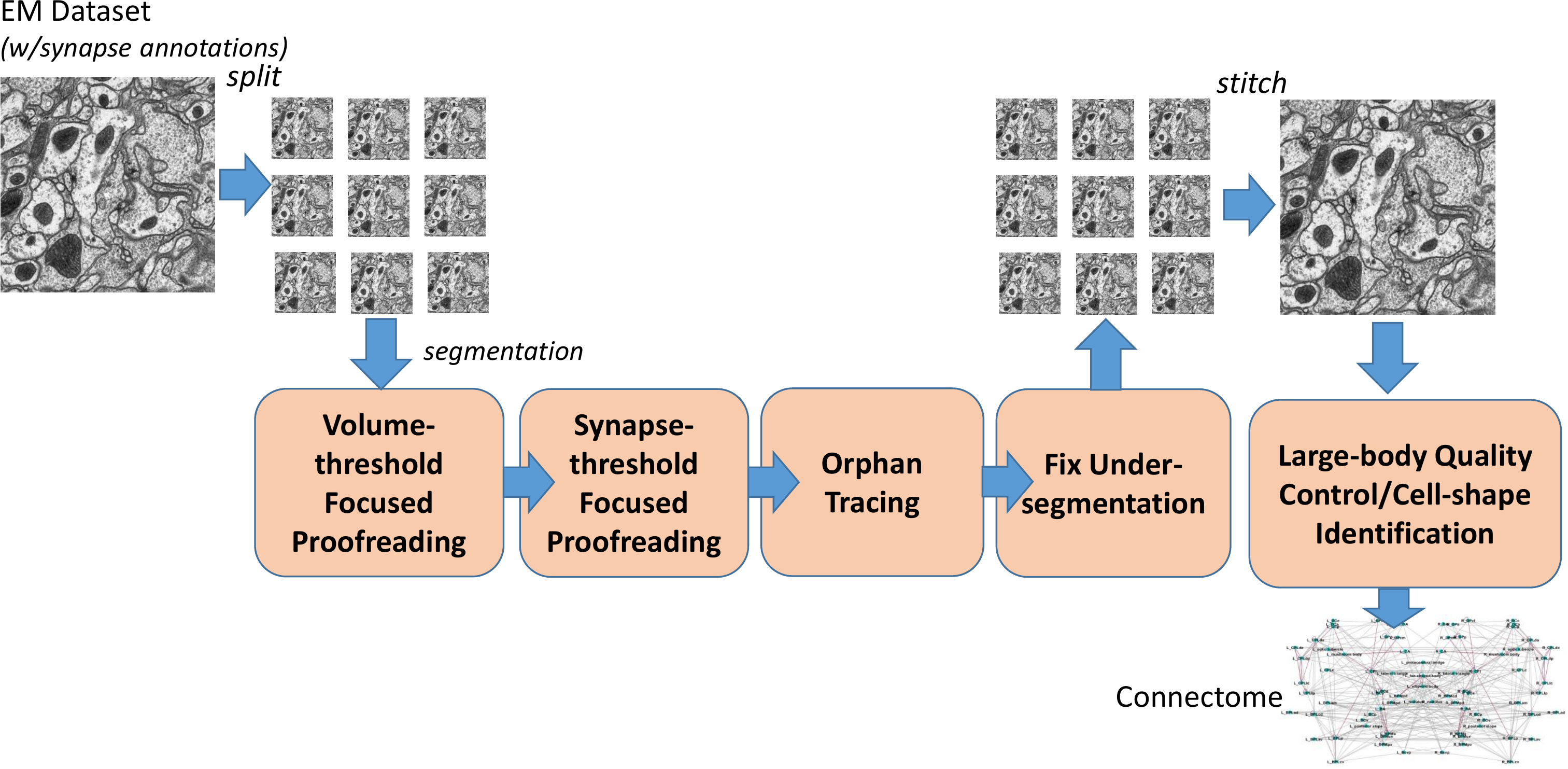}
\caption{\label{fig:segmentationflow} {\bf Segmentation-based proofreading methodology.}  The input is a subdivided dataset with synapse annotations already provided.}
\vspace{-1mm}
\end{figure}

\begin{figure}
\centering
\includegraphics[width=1.0\textwidth]{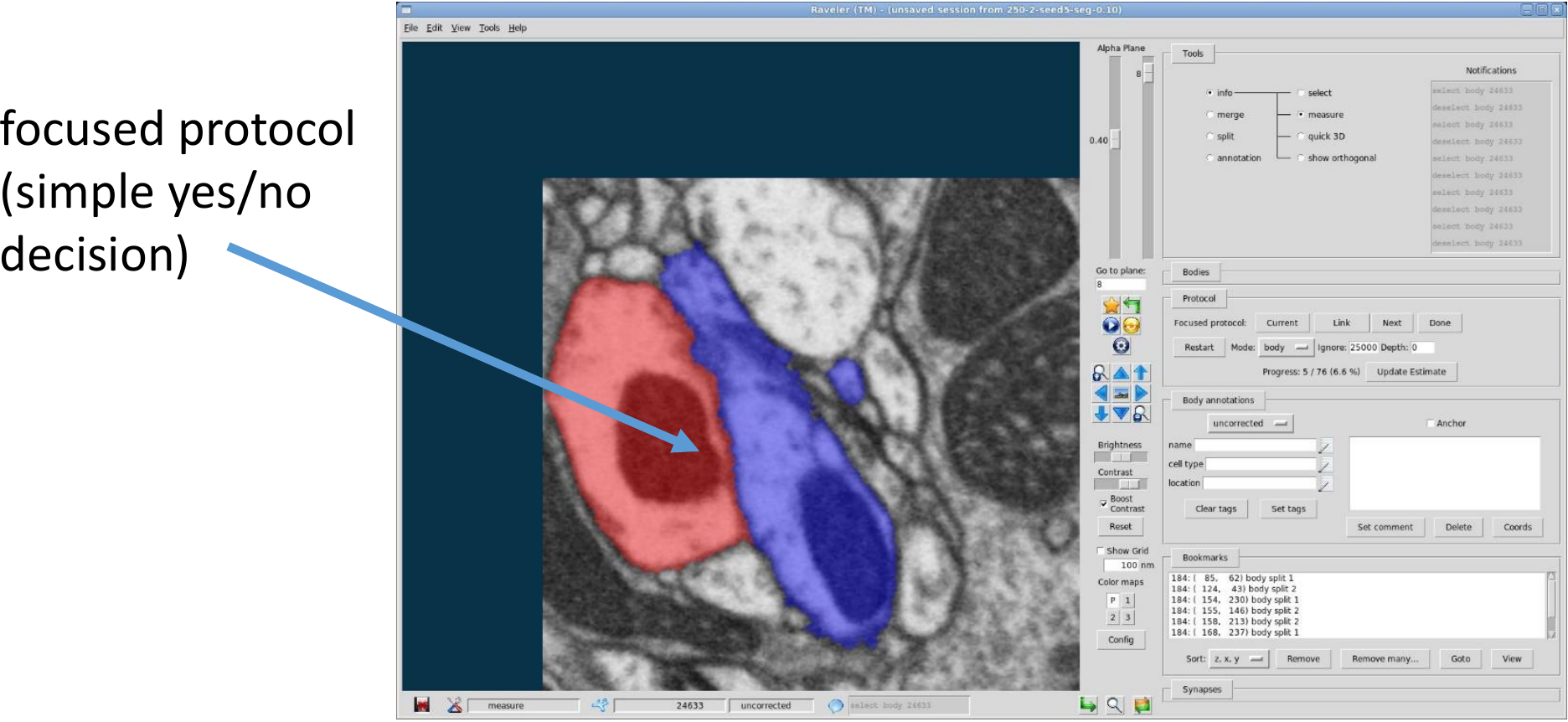}
\caption{\label{fig:focusedtool} Tool that implements focused protocol brings proofreaders to specific sites to render a yes/no decision.}
\vspace{-1mm}
\end{figure}

To address these concerns, we implemented the workflow shown
in Figure \ref{fig:segmentationflow}.  Because of the size of the volume, we divided
it into several subvolumes.  Each subvolume was separately proofread.
Three rounds of proofreading were performed: 1) volume-threshold focused proofreading, 2) synapse-threshold focused proofreading, and 3) orphan (small-body) tracing.  The first two rounds closely follow the algorithm in the preceding section but with different weighting strategies (we will investigate the advantage of doing volume-threshold before synapse-threshold in the experiments).  We also add some synapse connectivity constraints to eliminate unnecessary work.  For instance, in the synapse focused proofreading pass we ignore edges that would result in a rare autapse (a reflexive connection where a neuron drives itself).  Subsequent proofreading would uncover remaining orphan synapse processes.  Focused proofreading uses a special tool Raveler \cite{raveler} that highlights only the important edge, as shown in Figure \ref{fig:focusedtool}.  The goal is to reduce proofreading errors and variability between proofreaders of different experience levels.  The orphan (small-body) tracing is a quality control that has the proofreader examine disconnected segments that either contain synapses are of at least a certain size.  Therefore
if focused proofreading failed to connect certain regions, there
is some redundancy to ensure {\em important} areas are examined.

While proofreading these subvolumes, proofreaders note areas of false merging.  These areas were split in a separate pass after focused proofreading.  After proofreading each subvolume, we stitched them together to form a global segmentation.  In principle, it is possible to work off of an initial, global segmentation, but our workflow is computationally simple and easier to parallelize.  After stitching the region, additional quality controls and revisions are done on the reconstruction.  These quality controls involve looking for anomalies in connectivity and cell shape \cite{Ting14}.

\section{Experiments}
We implement the focused algorithms in a publicly available C++ tool called {\em NeuroProof}.  The thresholds and ordering strategy are primarily examined on a ground truth dataset from the Drosophila medulla volume produced from FIB-SEM imaging \cite{fib}.  FIB-SEM imaging produces volumes of near isotropic resolution, ideal for performing high-quality image segmentation.  We evaluate the consistency of proofreaders and reconstruction rates over several subvolumes in the medulla and compare to a previous reconstruction strategy.

The initial segmentation is generated using Ilastik \cite{ilastik} for voxel prediction and agglomeration algorithms introduced in \cite{Parag14} (and available in NeuroProof).  The synapses were
annotated before segmentation using the methodology defined in \cite{Plaza14Synapse}.  Proofreading and the focused proofreading protocols were performed with the open-source tool Raveler \cite{raveler}.

\subsection{Validation of Focused Proofreading}
In this section, we show that the proposed focused proofreading strategies are more efficient than other proofreading strategies.  We also validate some of our assumptions by showing the quality of the initial segmentation and predicted uncertainty.  The results are collected for a volume 500x500x500.  The difficulty of producing near-pixel perfect ground truth limits our ability to validate on more datasets.  We effectively increase our test set by running many of the experiments on $10$ random initial segmentations.

\begin{figure}
\centering
\includegraphics[width=0.8\textwidth]{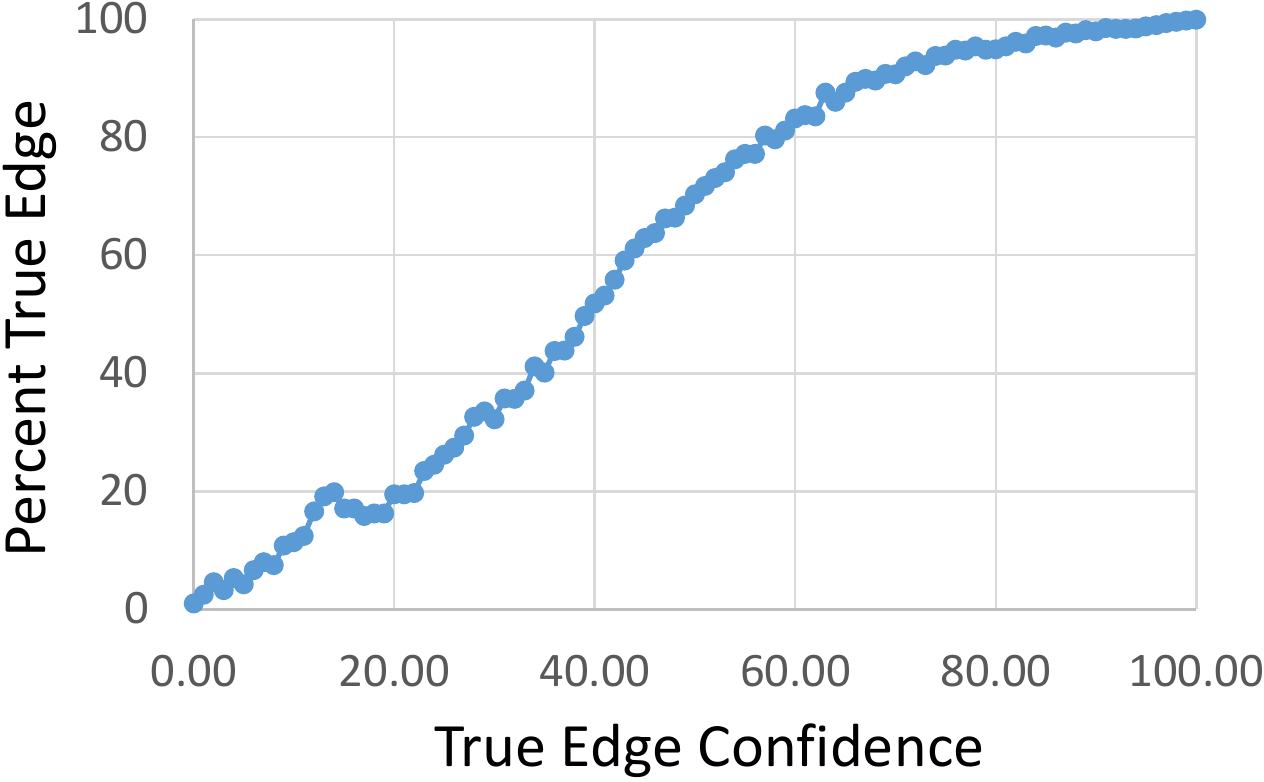}
\caption{\label{fig:uncertainty} Quality of the predicted edge confidences.  The plot show that the true edge prediction corresponds well with the actual true edge percentage.  At higher confidences the predictor tends to conservatively underestimate the number of true edges, which will result in more work.}
\vspace{-1mm}
\end{figure}

Before we show the effectiveness of the focused proofreading strategies
in this paper, we justify some of our assumptions.  Figure \ref{fig:uncertainty} shows the quality of the uncertainty estimation
produced by the segmentation classifier averaged over
$10$ runs.  Our algorithms depend on edge uncertainty predictions that closely reflect ground truth.  Our results show close correspondence between the predicted percentage of true edges and the number of actual true edges.  At higher true edge confidence, the distribution tends to under-estimate resulting in a conservative prediction of the actual percentage of true edges.  This could potentially result in more proofreading work since the segmentation appears more connected than it is.

\begin{figure}
\centering
\includegraphics[width=0.8\textwidth]{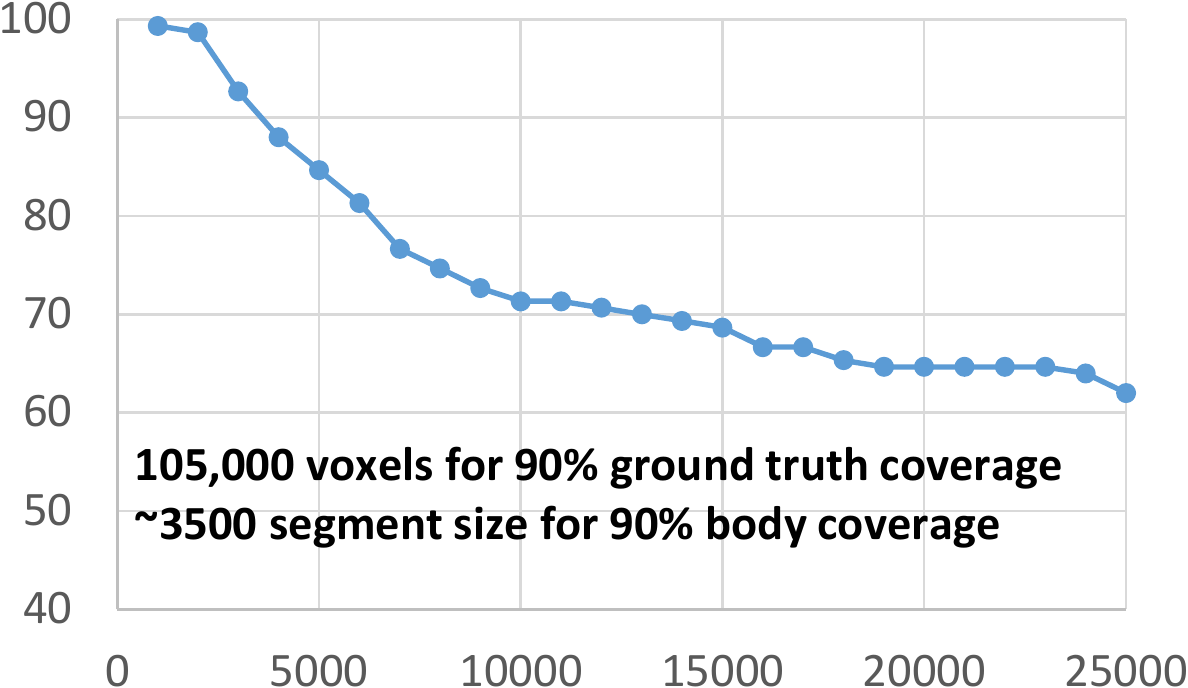}
\caption{\label{fig:coverage} Choosing focused proofreading
thresholds to achieve high coverage.  Based on ground truth, $|g|$ is determined to be around 105,000 voxels.  This plot shows the percentage of bodies adequately covered by segments at least of the given size $x$.  Around $x=3,500$ voxels, $90\%$ of the ground truth bodies are covered.  We choose this as the cut-off for the rest of our experiments.  $x$ need not equal $|s_{th}|$.}
\vspace{-1mm}
\end{figure}

We next justify our choice of parameters for our experiment.  Despite the formalisms describing reasonable segmentations and completeness, the choice of thresholds often comes down to heuristics and what looks reasonable.  Through inspection, we determine bodies of size $25000$ voxels to be important.  Figure \ref{fig:coverage} shows that bodies greater than $3500$ voxels need to be included to ensure that all important ground truth bodies are covered up to this $25000$ threshold.  (We consider important
bodies, the largest $N$ bodies that entail 90\% of the volume -- over $105,000$ voxels in this example.)  The final focused threshold value should
be probably be between the conservative $|s_{th}|=3500$, $|g|=7000$ and $|s_{th}|=25000$, $|g|\ge50000$ as defined in Equation \ref{eq:thres}.  
The chosen synapse threshold chosen is more straightforward.  Since $VI_{syn}$ effectively reduces the number of important points to only a few thousand, every synaptic point is important.  We set $|s_{th}|$ to $1$ and $|g|$ to $2$.

\begin{figure}
\centering
\includegraphics[width=1.0\textwidth]{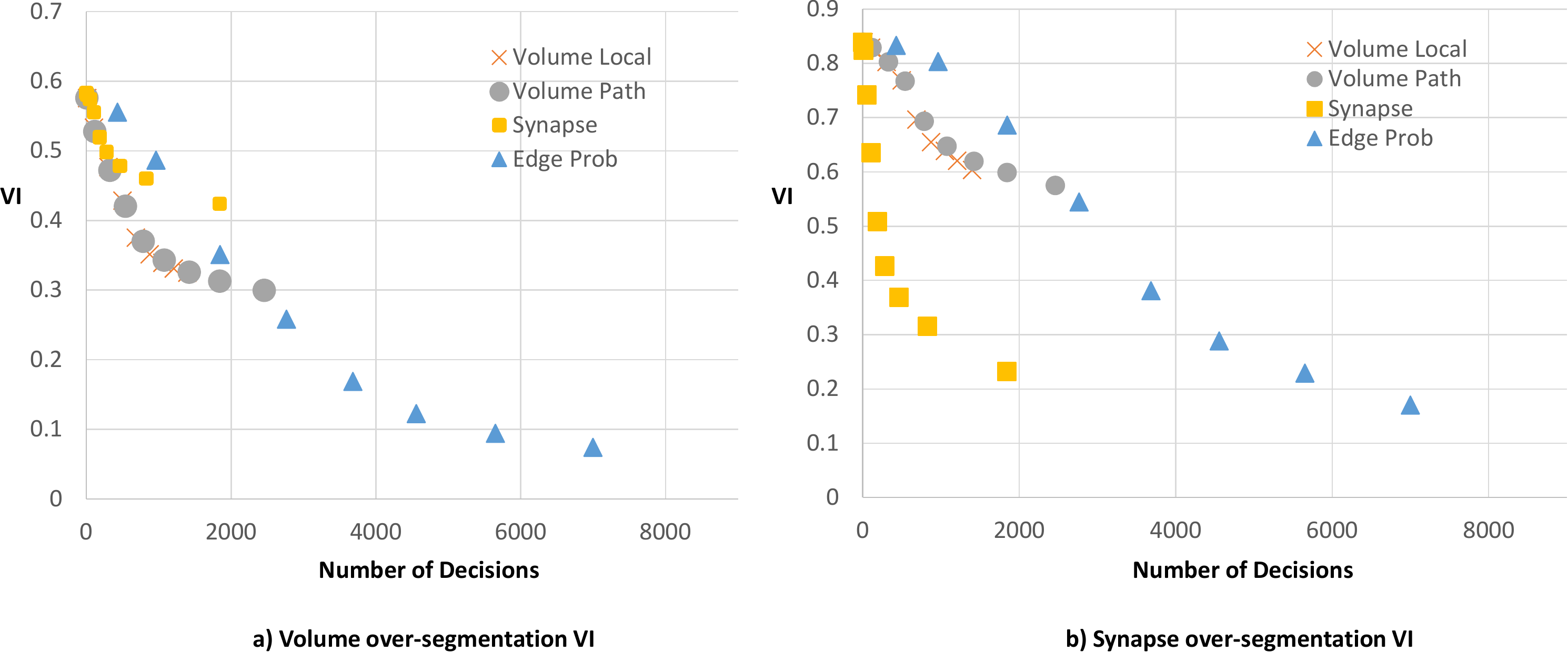}
\caption{\label{fig:vitrends} {\bf Improvements in VI over-segmentation similarity metric as a function of proofreader decisions.} Four ordering strategies are considered.  a) Shows slightly faster improvement using volume-based decision compared to using only edge confidence when considering volume VI.  b) Shows significantly faster improvement to synapse VI using synapse-based decisions.  This suggest that the edge probabilities are probably biased and more confident for larger bodies, not particularly useful when measuring connectivity that involves smaller processes.}
\vspace{-1mm}
\end{figure}

With adequate thresholds and good uncertainties, we now evaluate the trade-off between proofreading effort and proofreading quality using different focused proofreading heuristics in Figure \ref{fig:vitrends}.  We consider only the over-segmentation VI since the under-segmentation error is small and minimally impacted by our merge-only technique (under-segmentation proofreading will be discussed in the next section).  For these tests, proofreading effort is determined by automatically deciding on each edge presented.

In Figure \ref{fig:vitrends}a, we show volume VI trends.  Expectedly, the focused strategy that uses synapses for guidance does not do a good job improving the volume VI.  The two volume-guided focused strategies, volume-local and volume-path, do much better.  Volume-local only considers local bodies when making a decision.  Both perform similarly though volume-path achieves slightly lower VI by having a slightly longer cut-off.  We compared these approaches to a straightforward technique of using just edge probabilities.  The most confident false edges are chosen first.  This results in slightly worse, but comparable, results under $2000$ decisions.  However, more improvements are possible if one is willing to examine more edges.

Does this suggest that simple edge ordering is potentially sufficient?
First, focused proofreading explicitly chooses a stopping condition that trades-off errors.  The simplistic stopping condition for just using edge probability could result in a lot of unnecessary work.  Second, it appears that edges between big bodies (presumably where there is more boundary evidence) have more confidence.  This is apparently not the case for the smaller processes often important in tracing synapses.  The synapse VI plot in Figure \ref{fig:vitrends}b, shows that the synapse-guided mode is much better than all of the other techniques.
We note that random decision heuristics (not shown) perform much much worse than the above strategies.

\begin{figure}
\centering
\includegraphics[width=1.0\textwidth]{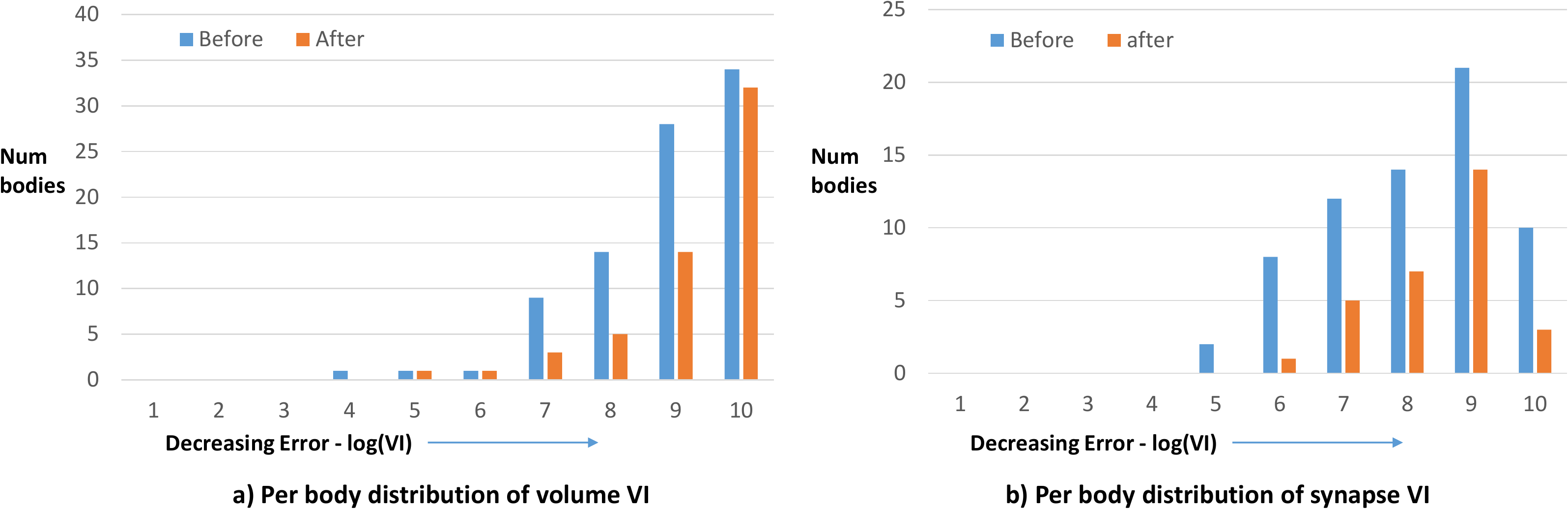}
\caption{\label{fig:vihist} Shows the improvement to the over-segmented VI metric after performing the workflow of volume plus synapse focused proofreading.  a) Shows changes to volume VI (log scale).  b) Shows changes to synapse VI (log scale).  In both cases, the distribution shifts to the right indicating improvement at the per body level.  The largest outliers are being improved.}
\vspace{-1mm}
\end{figure}

Figure \ref{fig:vihist} shows the distributions of over-segmented bodies before and after proofreading for one of our random runs.  We perform volume-path followed by synapse focused proofreading.  The $x$-axis is the $log_2$ of the VI.  Notice that for both volume and synapse VI, the distribution decisively shifts to the right after focused proofreading indicating that the worst bodies improved -- the goal of focused proofreading.

In general, the ordering of focused proofreading strategies seems to have a small effect
on the final similarity and effort required.  We notice a small, but statistically significant reduction in all $10$ trials in the number of examined edges of around $2\%$ when performing volume focused proofreading before synapse focused proofreading compared to trying synapse guidance first.  We also noticed a small increase in the average edge size of around $6\%$ when performing volume focused proofreading first.  We speculate that edges of larger size are generally easier for a proofreader to evaluate since there is more edge evidence.  For these reasons, we decide to perform volume proofreading first.

\subsection{Validation of Production Proofreading}
We deployed the focused proofreading strategies in a practical workflow to densely reconstruct seven columns of the Drosophila medulla optic lobe -- about $27,000$ cubic microns of EM data.  The reconstruction work described here (for synapse annotation times see \cite{Plaza14Synapse}) was primarily completed within 6 months.  To perform this work, we have a staff of 5-10
trained proofreaders.
  
We first compare the decisions between multiple proofreaders on a subvolume
and achieve agreement rates slightly over $98\%$.  The high consistency is motivation for applying only one proofreader per subvolume, followed by quality control that exploit biological priors and spot checking by senior biological experts.  While some specific pathways were revised by subsequent spot checking, we note that many motifs and connectivity patterns were unchanged.

\begin{table}
\centering
\begin{tabular}{|l|r|r|r|r|}\hline
Task & session hrs & working hrs & efficiency & microns/day \\ \hline \hline
focused proof & 3374 & 2934 & 87\% & 64 \\
synapse QC & 226 & 198 & 88\% & 956 \\
body split & 756 & 495 & 66\% & 286 \\
average & & & & {\bf 49.6} \\
\hline \end{tabular}
\caption{
\label{tab:7col}
Breakdown of proofreading effort in seven column medulla reconstruction (ignores time to annotate synapses
and downstream quality control). Focused proofreading includes body and synapse
VI, as well as, orphan tracing.  Synapse QC involves verifying and fixing some local connectivity anomalies observed in the data.  Body split is when under-segmentation is fixed.  All of these tasks are performed on subvolumes 125 microns in size.}
\vspace{-1mm}
\end{table}

We report the time to proofread the seven column medulla in Table \ref{tab:7col}.  Proofreading was performed over $216$ subvolumes each $125$ cubic microns and assigned randomly to the proofreaders.  The column {\tt session hrs} gives the amount time taken to complete the task. {\tt working hrs} gives the amount of time that the proofreader interacted with the proofreading tool (attempting to normalize for normal work distractions and circumstances where a proofreader needs to ask for help).  The ratio of working hours to session hours gives the {\tt efficiency}.  In general, 100\% efficiency is only possible for a robot.  Frustratingly challenging tasks tend to have a lower efficiency.  This could also be seen as a {\em frustration factor}.  {\tt microns/day} gives the rate of
cubic microns per session hour.

We show results for the following tasks: {\tt focused proofreading} (also includes the effort for orphan tracing), {\tt synapse QC}, and {\tt body split}.  {\tt synapse QC} has proofreaders review synaptic connections that seem suspicious, such as autapses.  {\tt body split} shows the time required to fix under-segmentation errors detected while focus proofreading.  Despite each subvolume requiring only 10s of splits, the task is time-consuming
and has reduced efficiency.  

Comparing our reconstruction efforts to the work in \cite{Nature13} is difficult
since the dataset in \cite{Nature13} was produced using serial section TEM imaging
resulting in a lower quality of segmentation.  The rate for proofreading subvolumes in \cite{Nature13} is around 10-20 microns per day (unpublished).  We believe the proofreading in this paper to be more comprehensive and results in a rate 3-5 times faster.  While much of the improvement likely stems from improved segmentation, our methodology
is much more focused, systematic, and less frustrating.

\section{Conclusions}
The time-consuming nature of EM reconstruction stymies our ability to understand larger, complex neurological systems.  This paper introduces a strategy called focused proofreading to greatly improve reconstruction speed allowing the analysis of much larger regions.  We demonstrated the effectiveness by reconstructing a complete connectome from a region of the Drosophila optic lobe, the largest such reconstruction ever performed.  The proposed workflow is amenable to large-scale, crowd-sourcing efforts.

This work is one of the first to focus on the quality of the uncertainty estimates of the segmentation engine, rather than just the resulting segmentation.  Future work should be directed at optimizing these confident intervals.  Furthermore, this work pioneers efforts at using biological priors and synaptic connectivity to guide proofreading process.  We believe exploiting more biological rules or priors can lead to great speedups.  Finally, this work emphasizes the need to decompose a complex task (proofreading) into a series of digestable decisions.  Additional work on improving visualization and making the task accessible to an even larger workforce should be explored.

{\small
\textbf{Acknowledgements:} We thank Zhiyuan Lu for sample preparation, Shan Xu and Harald Hess for FIB-SEM imaging; Pat Rivlin, Shin-ya Takemura, and
the FlyEM proofreading team (Roxanne Aniceto, Lei-Ann Chang, Shirley Lauchie, Mathew Saunders, Christopher Sigmund, Satoko Takemura, Julie Tran) for biological guidance and the reconstruction efforts; Donald Olbris for help with the proofreading tools; Louis Scheffer and Toufiq Parag for useful discussions and suggestions; and Ting Zhao for visualizations.}

\end{document}